\documentclass[preprint,aps,nofootinbib,11pt]{revtex4-1}
\usepackage{geometry}                
\usepackage{amsmath,amssymb,amsfonts,dcolumn,color,graphicx,graphics,latexsym,placeins,epsfig,tikz}
\usetikzlibrary{arrows,shapes}
\usepackage{subfigure,rotating,bm,mathrsfs}
\usepackage[title]{appendix}

\newcommand{\be}{\begin{equation}}
\newcommand{\ee}{\end{equation}}
\newcommand{\ba}{\begin{eqnarray}}
\newcommand{\ea}{\end{eqnarray}}

\usepackage[pagebackref=false, colorlinks=true]{hyperref}
\definecolor{redish}{rgb}{0.7,0.2,0.0}  
\definecolor{bluish}{rgb}{0.2,0.5,0.8}

\hypersetup{linkcolor=redish,          
                  citecolor=blue,        
                  filecolor=magenta,      
                  urlcolor=blue}          

\begin{document}

\title{\Large Analogue metric in a black bounce background}
\author{Kunal Pal, Kuntal Pal, Tapobrata Sarkar}
\address{
Department of Physics, Indian Institute of Technology Kanpur, \\ Kanpur 208016, India}
\email{kunalpal, kuntal, tapo@iitk.ac.in}

\bigskip

\begin{abstract}
The conventional approach of embedding an effective acoustic metric for
sound motion in a background flat Minkowski space-time, has recently been
extended to incorporate more general curved background metrics, which might
contain a black hole. Though the observational aspects of these
kinds of acoustics horizons, including the sonic shadow structure and quasi normal modes
have received significant attention in the literature, it leaves room for
discussions about embedding in more general classes of curved background space-times without optical horizons. 
Here we propose and study a new class of acoustic metrics that is embedded in a
black-bounce space-time, thereby giving a suitable tuneable system to
understand possible observational effects of the presence or absence of acoustic horizons. After
showing that the metric can represent five types of different effective
backgrounds for sound motion, including a novel ``acoustic wormhole - optical wormhole'' branch, we discuss how the distinctive features of
sonic shadows can appear even in the absence of any acoustic horizon due to the wormhole throat present in the acoustic metric.
\end{abstract}
\maketitle

\section{Introduction}
Einstein's general theory of relativity (GR) \cite{GR1} - \cite{GR4} remains the most successful theory of gravity till date and has been
extensively tested at different scales. In this context, over the last four decades, there has been strong interest in studying various 
analogue models of gravity, as these can mimic gravitational phenomena in tabletop experiments. Indeed, after the pioneering work of 
Unruh \cite{Unruh} (for an excellent review, see \cite{AnalogReview}), there have been seminal works in the literature that focus 
on experimental detection of gravitational effects via analogue gravity. 

Analogue gravity is essentially based on the idea that dynamical equations of excitations in physical systems can often be mapped to ones in
curved space-times. That is, these excitations ``see'' a curved space-time metric and this fact can then be used 
to obtain measurable effects in the laboratory that would then contain signatures of the metric itself. Physically, these would be reasonable
as long as the analogue space-time is stable, a topic that has been much discussed in the literature. Given that experiments at astrophysical
or cosmological scales present inherent difficulties, it is therefore natural that attempts at mimicking 
(i.e., engineering) appropriate curved space-times at such scales in the laboratory have been the focus of attention over the last
many years. Although the theoretical idea of analogue gravity is almost four decades old, major experimental progress have
happened only recently, and in the last decade, several such experiments have been reported, and promising results have been 
obtained. Till date, such experiments have been performed for example in the context of fluids, optical systems, ultra-cold atoms, 
etc. (for a sampling of the literature both on the theoretical and experimental aspects, 
see \cite{Visser1} - \cite{Ge2} and references therein. Some recent discussions on the validity and applicability of 
analogue models appear in \cite{Counter1},\cite{Counter2}). With sophisticated technologies that can be used to carry out
measurements with unprecedented accuracy rapidly becoming available, the hope is that analogue gravity experiments can be a tool 
to understanding deep aspects of the quantum nature of gravity, such as the nature of the black hole event horizon, and related quantum
fluctuations and entanglement.
  
Indeed, a significant part of studies on analogue gravity relate to black holes (BHs) -- singular solutions of Einstein's equations that possibly
hold the key to understanding strong gravity. Indeed, it is now believed that supermassive black holes exist in the central part of galaxies. 
The inherent importance of studying black holes lie in understanding one of the most intriguing unsolved questions regarding the fundamental nature
of matter, namely that of quantum mechanics in the regime of strong gravity (in other words, quantum gravity). In this backdrop, it is
natural that a large amount of existing literature seeks to understand so called ``acoustic'' black holes, a background formed, for example,
with moving fluids with the ``event horizon'' in such backgrounds being the location of the fluid where its speed exceeds that of sound. 

In a recent work, \cite{ge} extended the above line of work and 
showed how to embed analogue acoustic black holes in a curved background. This is in contrast to the
models studied earlier where analogue black holes were constructed in Minkowski space-time. In such a background, an excited state of
a fluid will ``see'' a more general acoustic metric than that of its flat space-time counterpart. This is important and 
interesting for two reasons. First, although often idealised by vacuum solutions of Einstein's equations, the background of 
astrophysical black holes are possible fluid systems that may arise due to effects such 
as accretion after tidal disruption processes. Hence it is more realistic to model analog astrophysical black holes in terms 
of fluids with a background curvature. Secondly, the extra freedom present with such curved backgrounds offer the 
interesting possibility of constructing exotic solutions that can possibly be mimicked in the laboratory. 

On the other hand, among the most intriguing observational aspects of ultracompact objects such as black holes, 
wormholes (WHs), naked singularities etc. that can bend light to form a photon ring, is the shadow of the given 
spacetime of interest. This research direction has seen a flurry of interest especially in the wake of the recent 
observation of the shadow of the supermassive object at the centre of our galaxy by the event horizon 
telescope \cite{EHT1,EHT2}. It is not hard to believe that such studies can also shed light on many important aspects of 
analogue gravity models and the respective acoustic metrics. In this regard, several works appear 
in the literature (see e.g., \cite{ling, ling2}) where the sonic shadow of the acoustic metric in a curved background 
is studied. On the other hand, in a real life laboratory setup, mimicking the actual light bending for sound waves 
in the presence of acoustic horizons can be problematic, mainly due to the fact that the acoustic horizon can be dynamically 
unstable due to the effect of the Hawking radiation present in the system (for extensive discussions see \cite{AnalogReview} 
and references therein). Hence, studying a system which can still offer the analogous effect of strong light bending 
even in the absence of the acoustic horizon is important, and understanding the differences of this, with that of
conventional acoustic black holes can offer valuable insights regarding fundamental aspects of analogue gravity.

In view of the above discussions, here we will take the background to be the 
recently constructed black bounce space-time of Simpson and Visser (SV) \cite{sv1}. 
The SV solution is a one-parameter family of space-times that interpolates between a deformed Schwarzschild
black hole and a traversable wormhole of the Morris-Thorne class. Using this as the background, here we construct an 
acoustic metric that can represent (1) a WH-WH system (2) a BH-WH system and (3) a BH-BH system with single 
and double horizons, depending on the SV parameter and the acoustic parameter that specifies the escape velocity from the background 
metric. In particular, we show that there are two extremely interesting possibilities that may arise. First, one can have 
sonic rings even in the absence of an acoustic event horizon. As already stated, this fact assumes importance given that 
such acoustic horizons may generically be dynamically unstable. Second, we show here that the acoustic SV background 
indicates that we can in principle distinguish between a WH-WH system and a BH-WH system by taking advantages of some of 
novel features of null geodesic motion in wormhole backgrounds.

This paper is organised as follows. The next section \ref{sec2} briefly reviews the necessary background material. 
In section \ref{sec3}, we construct the acoustic metric, whose structure is analysed in \ref{sec4}, where various
ranges of parameters are considered. Section \ref{sec5} deals with observational aspects of the solution, and
we conclude with some discussions in section \ref{sec6}. 

\section{Acoustic metric in curved background}
\label{sec2}
In this section we briefly review the construction of the acoustic metric from the Gross-Pitaevskii (GP)  
theory following \cite{ge}. We consider the propagation of a complex scalar field ($\Phi(x^{\mu})$) obeying the 
Gross-Pitaevskii equation on a static background metric of the form 
\begin{equation}\label{background}
ds^2_{bg}=g_{tt}(r)\text{d}t^{2}+g_{rr}(r)\text{d}r^{2}+r^{2}(\text{d}\theta^{2}+\sin^{2}\theta \text{d}\phi^{2})~.
\end{equation}
Substituting $\Phi(x^{\mu})=\sqrt{\rho(x^{\mu})}e^{\Theta(x^{\mu})}$ in the GP equation, 
then considering a perturbation around the background field $\Theta_0,~\rho_0$ and further 
working in the long wavelength limit, \cite{ge} derived a general form of the effective metric. The general
form of the acoustic metric is complicated and is in general not diagonal. Here, 
since we want to find out the acoustic geometry corresponding to the SV spacetime, it is sufficient 
to consider a simpler background static metric  such that $g_{rr}(r)g_{tt}(r)=-1$, and we will also assume that the $\theta$, 
$\phi$ components of the background four-velocity vanishes i.e., $v_{\theta}=v_{\phi}=0$. In that case, it 
is possible to render the acoustic metric in a convenient diagonal form, by making a change of the time coordinate as 
\begin{equation}
\text{d}t=\text{d}\tilde{t}+\frac{v_{t}v_{r}}{g_{tt}\big(c_{s}^{2}-v_{r}v^{r}\big)}\text{d}r~,
\end{equation}
where $c_{s}$ is the speed of sound and $v_{\mu}$ is the four-velocity of the  fluid. $c_{s}$ can be obtained by 
knowing the background value of the modulus of the complex scalar field, and by specifying the constants 
appearing in the GP theory. The phase of the complex scalar field determines the components of the 
four-velocity by the relations $v_t=-\partial_t \Theta_0$ and $v_i=-\partial_i \Theta_0$ with $i=r,\theta,\phi$.  
The conditions  $v_{\theta}=0$ and $v_{\phi}=0$ mean that the background phase $\Theta_0$ is independent 
of the angular coordinates $\theta,\phi$.
Furthermore, we shall work in the critical temperature of the GP theory so that the normalisation condition satisfied by the 
background four-velocity of the fluid is $v_{\mu}v^{\mu}=-2c_{s}^{2}$.

Following all the above steps, the acoustic metric derived \cite{ge} in terms of  the coordinates 
$\tilde{t},r,\theta,\phi$  is given by
\begin{equation}
ds^2_{ac}=c_{s}\sqrt{c_{s}^{2}-v_{\mu}v^{\mu}}\Bigg[\frac{c_{s}^{2}-v_{r}v^{r}}{c_{s}^{2}-v_{\mu}v^{\mu}}g_{tt}
\text{d}\tilde{t}^{2}+\frac{c_{s}^{2}}{c_{s}^{2}-v_{r}v^{r}}g_{rr}\text{d}r^{2}+r^{2}\big(\text{d}\theta^{2}+\sin^{2}\theta 
\text{d}\phi^{2}\big)\Bigg]~.
\end{equation}	
Thus  the 
time coordinates $\tilde{t}$ appearing in the acoustic line element $ds^{2}_{ac}$ is different from the time 
coordinate $t$ appearing in the background line element. 

Now we re-scale the velocity vector as $v_{\mu}v^{\mu}=2c_{s}^{2}\tilde{v}_{\mu}\tilde{v}^{\mu}$, such that the new 
one satisfies the usual normalisation condition $\tilde{v}_{\mu}\tilde{v}^{\mu}=-1$.  Substituting this in the 
acoustic line element, and removing all the tildes for notational convenience, and further 
denoting $g_{tt}(r)=f(r)=g_{rr}(r)^{-1}$, we obtain the following form for the acoustic metric
\begin{equation}\label{acoustic}
ds^2_{ac}=\sqrt{3}c_{s}^{2}\Bigg[\big(1-2v_{r}v^{r}\big)f(r)\text{d}t^{2}+\frac{1}{f(r)
\big(1-2v_{r}v^{r}\big)}\text{d}r^{2}+r^{2}\big(\text{d}\theta^{2}+\sin^{2}\theta \text{d}\phi^{2}\big)\Bigg]~.
\end{equation}	
From now on we will put $\sqrt{3}c_{s}^{2}=1$. Given a static background metric of the form in Eq. (\ref{background}) 
obeying $g_{rr}(r)g_{tt}(r)=-1$ i.e. with specified functional form of $f(r)$ we can easily obtain the corresponding 
acoustic metric using Eq. (\ref{acoustic}). This form of acoustic metric has attracted a lot of recent attention due 
to its simple analytical form and potential experimental applications, and has been explored recently in \cite{ling}-\cite{yuan}.
Notably, most of the above cited recent works explore the acoustic metric in black hole backgrounds, 
with stationary and slowly rotating versions considered in \cite{viera}. This leads to the natural question 
whether we can generalise this metrics for more generic backgrounds, like wormhole geometries or non singular geometries, 
and if so, what special features those acoustic metric might show. 

This is what we set out to do in the next sections and our aim  will be to construct the corresponding 
acoustic metric and study its properties, when the background is the recently proposed one parameter 
generalisation of the Schwarzschild metric, namely the SV \cite{sv1} geometry.

\section{Acoustic metric in the black-bounce background}
\label{sec3}
As a background metric we choose to work with the black-bounce space-time introduced by SV, 
that can represent a regular black hole (BH) or a (one-way or two-way) traversable wormhole (WH), depending on the 
parameter range of interest. The SV solution is given by
\begin{equation}\label{RJNW}
ds^2_{bc}=-\Big(1-\frac{2M}{\sqrt{r^2+\beta^2}}\Big) \text{d}t^2 + \Big(1-\frac{2M}{\sqrt{r^2+\beta^2}}\Big)^{-1} 
\text{d}r^2 + \Big(r^2+\beta^2\Big) \text{d}\Omega^2~.
\end{equation}
Here $\beta>0$ is a constant and the range of the $r$ coordinate is $-\infty$ to $\infty$. 
For different values of $\beta$ (in comparison to the mass $M$), the above metric may represent the following : 
(1) For $\beta>2M$, the SV metric represents a two-way traversable wormhole, 
(2) when $\beta=2M$, it describes a one-way wormhole, and finally, 
(3) for $\beta<2M$ it is a regular black hole. The introduction of the non zero constant $\beta$ 
helps to remove the central singularity of the Schwarzschild space-time. In the last case, 
the regular black hole has two horizons at the locations $r_{\pm}=\pm\sqrt{(2M)^{2}-\beta^{2}}$.

Let us now  consider the motion of a  relativistic fluid moving in this background.  
Following \cite{ling}, we consider the radial velocity of the fluid to be of the form 
$v_{r}=\sqrt{\frac{2M}{\sqrt{r^2+\beta^2}}\xi}$  such that the fluid is not trapped by the background geometry. 
This can be ensured by choosing the  radial velocity to be greater than the escape velocity i.e., 
by taking the factor $\xi \geq 1$. As mentioned before,  working in the critical temperature regime of 
Gross-Pitaevskii theory, and after the rescaling of the four velocity, we arrive at the following 
acoustic metric as ``seen'' by sonic excitations,
\begin{equation}\label{asv}
ds^2_{ac}=\sqrt{3}c_{s}^2\Big[-F(r) \text{d}t^2 + G(r)^{-1} \text{d}r^2 + \Big(r^2+\beta^2\Big) \text{d}\Omega^2\Big]~,
\end{equation}
where we have defined
\begin{equation}\label{grrgtt}
F(r)=G(r)=\Bigg(1-\frac{2M}{\sqrt{r^2+\beta^2}}\Bigg)\Bigg[1-\frac{2M\xi}{\sqrt{r^2+\beta^2}}
\Bigg(1-\frac{2M}{\sqrt{r^2+\beta^2}}\Bigg)\Bigg]~.
\end{equation}
It is easy to check that this metric posesses all the well defined limits that has appeared in the literature, 
namely in the limit $\xi \rightarrow 0$, it reduces to the original back ground SV metric. 
On the other hand, in the limit $\beta \rightarrow 0$, it gives us the acoustic metric in the 
Schwarzschild background discussed recently in \cite{ling2}.
	
\section{Structure of the acoustic metric}
\label{sec4}
In this section, we discuss details the nature of the objects that the acoustic metric can represent, depending 
on the different ranges of the  parameter values involved. To analyse this more clearly, we perform a simple 
coordinate transformation $r^2+\beta^2=k^2$, such that the area of the two sphere part is now given by 
$4\pi k^2$. And the (radial) $g_{kk}$ term now reads 
\begin{equation}
g_{kk}=\frac{k^2}{(k^2-\beta^2)\big(1-\frac{2M}{k}\big)\Big(1-\frac{2M\xi}{k}\big(1-\frac{2M}{k}\big)\Big)}~.
\end{equation}
The last factor in the denominator gives the polynomial $k^2-2M\xi k +4M\xi^2 $, and hence it can be seen that 
the denominator has four positive roots and their interplay will determine the nature of the acoustic metric. 
The four roots are
\begin{equation}
k_{1}=\beta~,~~~k_{2}=2M~,~~\text{and}~~k_{3,4}=M\xi \pm M\sqrt{\xi^2-4\xi}~.
\end{equation}
In  comparison with the original SV metric (when written in the $k$ coordinate), we can see that the first two 
roots $k_{1}, k_{2}$ have their origin in the background SV metric, whereas the novel contribution of the fluid velocity 
in the  acoustic metric (through the parameter $\xi$) is manifested in the last two solutions $k_{3}, k_{4}$. 
In order to make these roots real, we can see that the allowed range of parameter $\xi$ has to be  $\xi \geq 4$ 
and for $\xi=4$, these two roots coincide, making the solution extremal. Also note that when 
$k_{3}, k_{4}$ are real, they are always greater than $k_{2}$.

Considering  the motion of a null particle ($ds^2=0$) we  have 
\begin{equation}\label{lightcone}
\frac{\text{d}r}{\text{d}t}=F(r)=\Bigg(1-\frac{2M}{\sqrt{r^2+\beta^2}}\Bigg)\Bigg[1-\frac{2M\xi}
{\sqrt{r^2+\beta^2}}\Bigg(1-\frac{2M}{\sqrt{r^2+\beta^2}}\Bigg)\Bigg]~.
\end{equation}
From this light-cone structure, we can infer different cases depending on the value of $\beta$. 
The solution of the equations
\begin{equation}\label{lightcones}
1-\frac{2M}{\sqrt{r^2+\beta^2}}=0~,~~1-\frac{2M\xi}{\sqrt{r^2+\beta^2}}\Big(1-\frac{2M}{\sqrt{r^2+\beta^2}}\Big)=0~,
\end{equation}
will give rise to none, one or many horizons, as these are the positions where the light cones flip. 
For later purposes, we note the solutions to be (only those are positive for the parameter ranges 
and hence on ``our side'' of the universe)
\begin{equation}\label{acostichorizon}
\begin{split}
r_{1}=\sqrt{4 M^2-\beta^2}~,~~ r_{2}=\sqrt{2 \left(M^2 (\xi -2) \xi-\sqrt{M^4 (\xi -4) \xi ^3} \right)-\beta^2}~,~~\\
~~\text{and}~~
r_{3}=\sqrt{2 \left(\sqrt{M^4 (\xi -4) \xi ^3}+M^2 (\xi -2) \xi \right)-\beta^2}~.
\end{split}
\end{equation}
If we take  $\xi \geq 4$, then both the solutions $r_{3}$ and $r_{2}$ are greater than $r_{1}$. 
 The optimal ranges of $\beta$ upon which the structure of the metric depends can be easily obtained by solving 
Eq. (\ref{lightcones}) for $r=0$, which are identical to that of $k_{2}, k_{3}, k_{4}$.

\subsection{Case 1: $\beta>$ $k_{4}, \xi \geq 4$ : Acoustic wormhole, Optical wormhole} 
In this parameter range, for sound waves, the metric behaves like a two way traversable wormhole with a 
throat at $r=0$, and for light rays (which move in the background metric), the system behaves like a wormhole. 
We will call this a WH-WH system. 
This can be seen from Eq. (\ref{lightcone}), that since $\frac{\text{d}r}{\text{d}t}$  
is never zero with this range of parameters, the absence of a horizon in the acoustic metric is guaranteed. It can also be 
seen that in this case, the acoustic metric consists of two copies of same metric glued at $k=\beta$ ( in the 
original coordinates this location is at $r=0$) since the metric can be extended from $r=-\infty$ to $r=0$ to 
$r=\infty$.\footnote{In an experimental setup, we cannot have an acoustic wormhole that connects two asymptotically 
flat space-times, but here we are only interested in ``our side'' of the wormhole, where the acoustic analogue of 
the WH throat may leave its imprint. See \cite{Nandi} for a detailed discussions.}
The traversibility can be confirmed by checking the so called `flaring out' condition for a wormhole, 
which ensures that the throat is the location of minimum area, and establishes the absence of a horizon in this 
space-time \cite{simpsonthesis}. Mathematically, this demands that at the throat $\mathbf{R}^{\prime}(r=r_{0})=0, 
\mathbf{R}^{\prime\prime}(r=r_{0})>0$ with $\mathbf{R}(r)$ being the two sphere part of the metric. 
It is also crucial to check that the geometry is non singular everywhere, which can be confirmed by calculating the 
Ricci and the Kretschmann scalar. The Ricci scalar corresponding to the metric in Eq. (\ref{asv}) is 
\begin{equation}
\begin{split}
R(r)=\frac{1}{\big(r^{2}+\beta^{2}\big)^{7/2}}\Big[16M^{3}r^{2}\xi+\beta^{4}\Big(6M\big(1+\xi\big)-2\sqrt{r^{2}+\beta^{2}}\Big)
+\beta^{2}\Big(6Mr^{2}\big(1+\xi\big)+8M^3\xi\\
-16M^2\sqrt{r^{2}+\beta^{2}}-2r^2\sqrt{r^{2}+\beta^{2}}\Big)\Big]~.
\end{split}
\end{equation}
It can be easily checked that as $r \rightarrow 0$, the Ricci scalar takes a constant value. 
A similar conclusion is valid for the Kretschmann scalar as well.

This particular branch  of the acoustic metric offers novel features of
sound motion which, to the best of our knowledge, have not been appeared in
the literature before. As we shall describe below, here we can
observe strong bending of sound and consequently sonic rings even in the
absence of any acoustic event horizon - which is a specialty of photon
motion in a wormhole background. In view of the generic dynamical instability of the
acoustic event horizons \cite{AnalogReview}, this regular WH-WH system can offer an
interesting `stable' system to experimentally observe the related
observational signatures. As a caveat, we add that admittedly, the construction of wormholes
in general relativity require so called `exotic matters' that might violate the standard
energy conditions, but these can be realised in many modified gravity setups
with regular matter sources.

\subsection{Case 2: $k_{3} < \beta < k_{4}, \xi \geq 4$ : Acoustic black hole (one horizon), Optical wormhole}

This is another interesting situation, where the metric posesses an outer acoustic horizon at $r=r_{3}$ so that  
the acoustic metric describes an acoustic black hole with a space-like throat (though this term should not be 
strictly used for sound waves) at $r=0$ that is only one way traversable (due to the presence of the 
horizon). This is seen from the fact that the light cone flips  at $r=r_{3}$. Crucially, in this 
case the space-time metric is still a wormhole (two way traversable) with a time-like throat at $k=\beta$ ($r=0$) 
since by the imposed conditions above, $\beta > k_{2}$ is always satisfied. We call this a BH-WH system. 

Here the appearance of the acoustic horizon in the absence of any
background event horizon signifies the fact that the sound cone can flip even
though the light cone does not. Hence every optical horizon is a acoustic
horizon but the reverse is not true.  This type of regular acoustic
geometry (BH-WH system) is also a novel addition to the existing
literature, and offers a different sonic shadow structure to that of the WH-WH
system of case 1, depending on the source of sound present. It is
well known that if the source is present on the other side of the
throat compared to the observer, the wormhole can cast qualitatively
different shadow than that of a black hole. In an entirely similar way, here we
can in principle distinguish the two different systems (WH-WH and BH-WH) by
observing their corresponding sonic rings.

\subsection{Case 3: $k_{2} < \beta < k_{3}(<k_{4}), \xi \geq 4$ : Acoustic black hole (two horizons), Optical wormhole}

In this case, the metric possesses both inner and outer horizons at $r_{3}$, and $ r_{2}$, since both of them are real when 
the above restrictions are imposed. Thus, the acoustic metric represents a multi-horizon acoustic black hole with a 
time-like throat at $r=0$. This is in contrast with case 2 above, where the throat  was space-like. But again (like case 2) 
the optical geometry is a two way traversable wormhole since the condition $\beta> k_{2}$ is still satisfied here.

\subsection{Case 4: $\beta<k_{2}, \xi \geq 4$ : Acoustic black hole, Optical black hole}

Here, $\beta$ is less than $k_{2}$, hence the acoustic metric is a acoustic black hole with three horizons (one coming 
from the original background metric and the other two are acoustic horizons). Also, the background is 
a regular black hole with a single horizon at $k=2M$ with a space-like throat at $r=0$.
This case is qualitatively similar to the one reported in \cite{ge, ling}, where the background geometry was taken to be the
Schwarzschild solution.

\subsection{Case-5: $\xi <4$ Acoustic and optical geometry qualitatively same}

Finally, in this case none of the solutions $r_{2}, r_{3}$ are real and the acoustic metric has the same 
structure as that of the background SV metric (though they have different coordinate dependence), 
and will represent a black hole or traversable wormhole depending on the values of the SV parameter $\beta$. 

\section{Observational aspects of the black-bounce acoustic metric}
\label{sec5}

One of the most studied aspects of a space-time metric containing a black hole or a wormhole is its 
shadow structure, which is defined as the region of the observer's sky which is left dark. This aspect 
of BHs and WHs is extremely well studied in the literature, and its importance has increased significantly after the 
recent observation of the shadow of the compact object at the centre our galaxy by the Event Horizon Telescope. 
In the traditional study of the shadow of a compact object (with or without a horizon), 
an important aspect is the behaviour of null geodesics, which can lead to interesting phenomena, 
like the formation of Einstein rings. 

For the spherically symmetric and  static metrics that we have considered, the study of null geodesics are simple, 
due to the presence of integrals of motions, and hence can be treated analytically. Here our 
main goal will be to study the observational aspects of the acoustic metric in black-bounce space-times 
derived above by the standard assumption that the sound waves move in null geodesics in this acoustic metric.
This is in the same sense as light waves moving in a space-time metric, such that the standard approach to solving this
problem involving the effective potential (encountered by a photon) can be implemented. In  
particular, our emphasis will be on the availability and the study of the analogue of the `photon sphere' or the  
`anti-photon sphere', which are the unstable (or stable) circular orbits of photons (here sound waves) 
that may give rise to \textit{acoustic shadow} for some listener far away from the acoustic black hole (or wormhole). 

The conventional analytical approach to study the motion of the null geodesics is to employ the Hamilton-Jacobi 
equation given by
\begin{equation}
\frac{\partial S}{\partial\lambda}+H=0~,~ H=\frac{1}{2}g_{\mu\nu}p^{\mu}p^{\nu}~.
\end{equation}
Here, $H$ is the Hamiltonian of the particle, $S$ is the Jacobi action, $\lambda$ is an 
affine parameter that parametrises a geodesic, and  $p^{\mu}$ is the $4$-momentum vector  
defined by $p_{\mu}=\frac{\partial S}{\partial x^{\mu}}$. Since  the acoustic metric  is static and 
spherically symmetric, there are two constants of motion associated with the particle motion, 
namely the energy $E=-p_{t}$ and the angular momentum $L=-p_{\phi}$. Then assuming  a separable form of 
the Jacobi action, we can write down the effective potential for the motion of the null rays in the 
radial direction to be \cite{Sau}
\begin{equation}\label{veff}
V_{eff}(r)=\mathcal{G}_{tt}\frac{L^2}{\textbf{R}(r)}~,
\end{equation}
where ${\mathcal{G}_{tt}}$ is  the temporal component of the acoustic metric, and $\textbf{R}(r)$ is its 
two sphere part. Now the \textit{photon sphere} (or the \textit{anti-photon sphere}), 
which determines the boundary of the shadow is given by the extrema of the effective potential, 
and our study in this case reduces to analysing the resulting effective potential for this metric.

Substituting $\mathcal{G}_{tt}$ from Eq. (\ref{grrgtt}) in  Eq. (\ref{veff})   
we have  the  expression for the effective potential to be
\begin{equation}
V_{eff}(r)=\frac{L^2}{\beta^2+r^2} \left(1-\frac{2 M}{\sqrt{\beta^2+r^2}}\right) \left(1-\frac{2 M \xi  
\left(1-\frac{2 M}{\sqrt{\beta^2+r^2}}\right)}{\sqrt{\beta^2+r^2}}\right)~.
\end{equation}
Now imposing the condition $V^{\prime}_{eff}(r)=0$, we can read off the extrema of the effective potential. 
In this case, these are given by the solutions of the equation 
\begin{equation}\label{ps}
r^2 \sqrt{\beta^2+r^2}-20 M^3 \xi +16 M^2 \xi  \sqrt{\beta^2+r^2}-3 M (\xi +1) r^2+\beta^2\Big(\sqrt{\beta^2+r^2}-3 M (\xi +1)\Big)=0~.
\end{equation}
A trivial solution is at $r=0$, which will of interest later in the wormhole case.  
The analytical expressions for the other solutions of this equation are complicated, and we will not write them 
here for brevity. Rather we plot the effective potential to directly obtain the corresponding locations 
of the maxima (corresponding to the unstable photon orbits, i.e., the  photon sphere) 
and minima (stable photon orbits, i.e., the anti-photon sphere) for given values of the 
parameters. It is important to note that when the acoustic metric represents a WH, 
apart from the photon spheres that are outside the throat (solution of Eq. (\ref{ps}) 
with maximum of the effective potential), the throat itself can act as a location of the unstable 
light rays, and hence can give rise to an additional photon sphere. For details of this and 
related issues see \cite{shaikh2}. Now we shall consider the different cases that can appear.

\subsection{Case-1: Acoustic wormhole ($\beta>$ $k_{4}, \xi \geq 4$) } 
In this case, the metric represents an acoustic wormhole, and the condition $\beta>k_{4}$ 
imposes a restriction on the possible values that the SV parameter can take in this range, 
which in turn gives two types of effective potentials. Below we discuss different possible 
scenarios depending on the value of $\beta$ for a particular choice of $M$ and $\xi$, 
taken for the purpose of illustration as $M=1, \xi=4.1 $. From Eq. (\ref{ps}), we see that for these values 
of the parameters, this equation can have either one positive real solution or no real root at all.

\begin{figure}[h!]
\begin{minipage}[b]{0.4\linewidth}
\centering
\includegraphics[width=2.7in,height=2.0in]{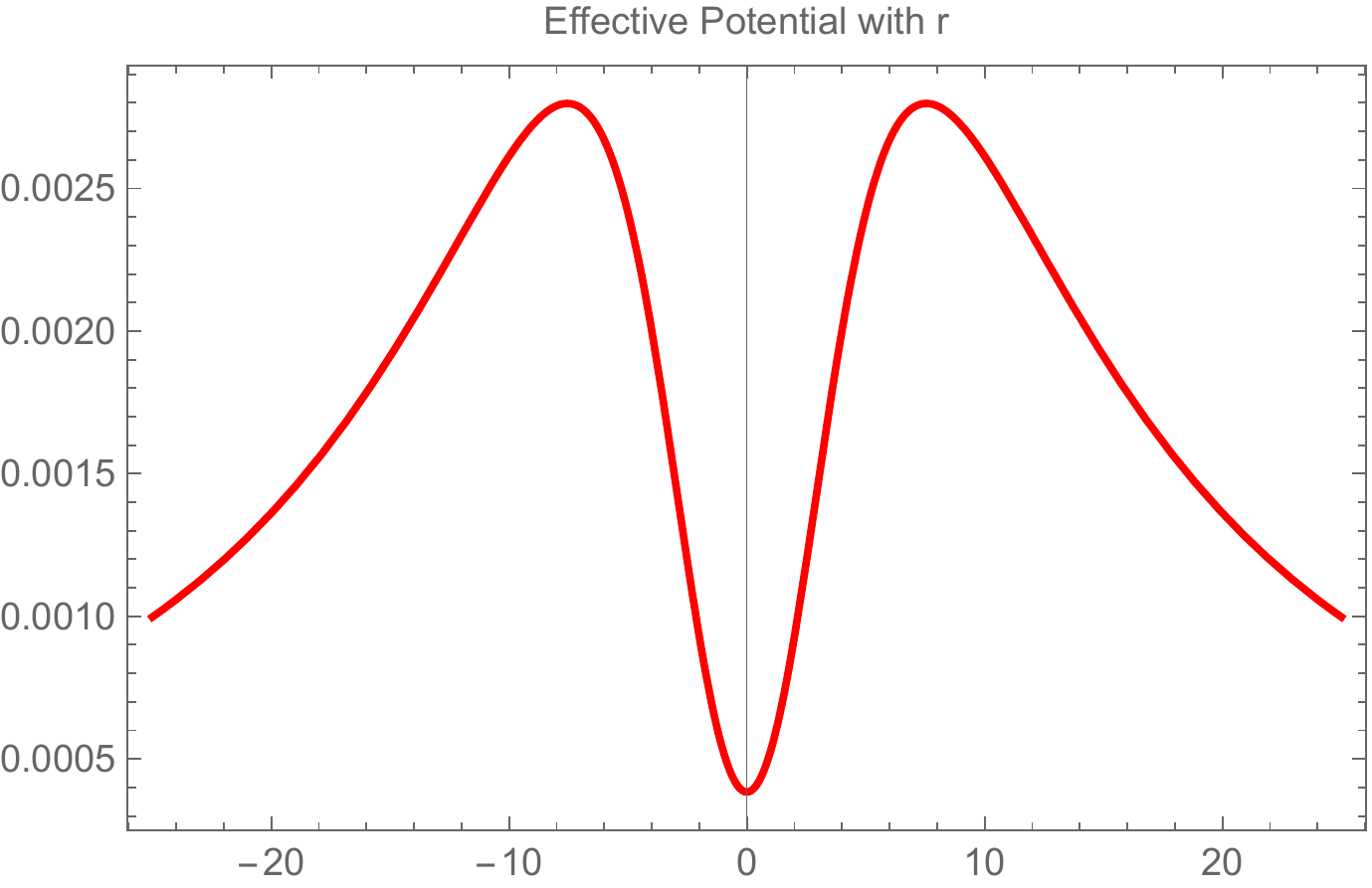}
\caption{Variation of the effective potential with $r$ for $\beta=5$, $M=1$, $L=1$, and $ \xi=4.1$.}
\label{fig:CASE1.1}
\end{minipage}
\hspace{0.2cm}
\begin{minipage}[b]{0.4\linewidth}
\centering
\includegraphics[width=2.7in,height=2.0in]{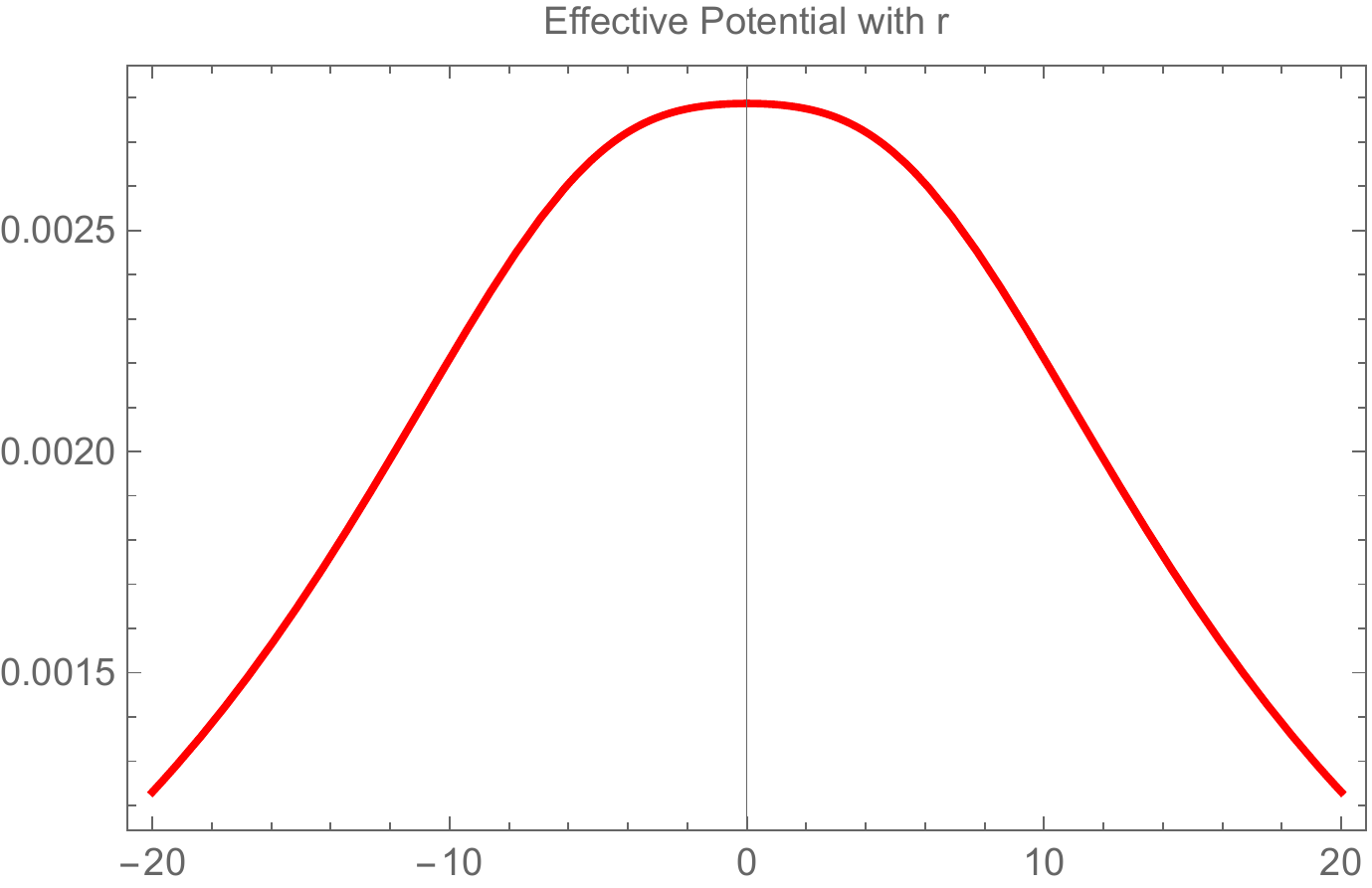}
\caption{Variation of the effective potential with $r$ for $\beta=9.5$, $M=1$, $L=1$, and  $\xi=4.1$.}
\label{fig:CASE1.2}
\end{minipage}
\end{figure}

\subsubsection{Case 1.1 : One photon sphere per side}
For the choice of parameters $M=1, \xi=4.1 $ and $L=1$, we find that $k_4=4.74$, so that we  
are restricted to take the value of $\beta$ to be above this limit. As can be directly seen from the plot of the 
effective potential shown in Fig. \ref{fig:CASE1.1}, for values of $\beta>k_4=4.74$, 
upto a critical value, the null rays encounter only one photon sphere per side of the throat. 
Note that the numerical choices made here are for illustration only. For example, 
for different values of $L$, the profile of the effective potential remains qualitatively same, 
and only its magnitude is changed.

\subsubsection{Case 1.2: No photon sphere outside the throat}
If the value of the SV parameter $\beta$ goes above a certain limit, we observe an 
interesting feature of the effective potential that is typical of many wormhole solutions. Here, 
none of the roots of the Eq. (\ref{ps}) are real, and instead the effective potential is maximum at $r=0$ 
indicating that the throat itself is a location of the unstable photon orbits. We have shown this case for 
$\beta=9.5$ for $M=1, \xi=4.1$ and $L=1$ in Fig. \ref{fig:CASE1.2}.

In both these cases, the sonic ring is formed in an acoustic wormhole metric where the acoustic horizon is absent. 
This is a special feature of many wormhole metrics. Note however that though the above two cases represents WH-WH systems, 
the strong bending formula and associated observables are different for case (1.1) and case (1.2), which indicates
an important distinction between these two cases (see \cite{shaikh1} for details).

\subsection{Case 2: Single horizon  Acoustic black hole ($k_{3} < \beta < k_{4}, \xi \geq 4$)} 
\begin{figure}[h!]
\begin{minipage}[b]{0.4\linewidth}
\centering
\includegraphics[width=2.7in,height=2.4in]{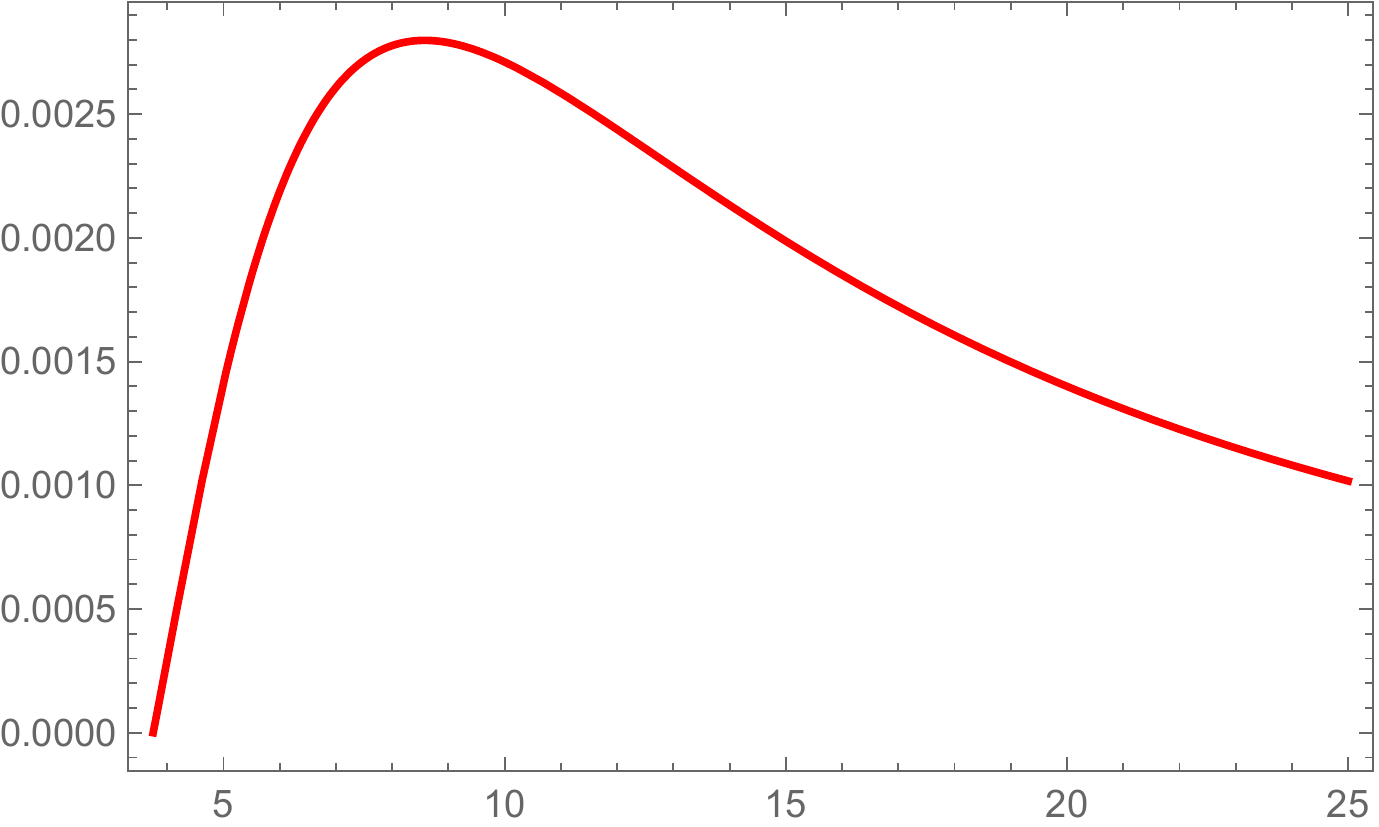}
\caption{Variation   of the effective potential with $r$ for $\beta=2.9$ for $M=1, L=1$ and $ \xi=4.1$.}
\label{fig:CASE2}
\end{minipage}
\hspace{0.2cm}
\begin{minipage}[b]{0.4\linewidth}
\centering
\includegraphics[width=2.6in,height=2.4in]{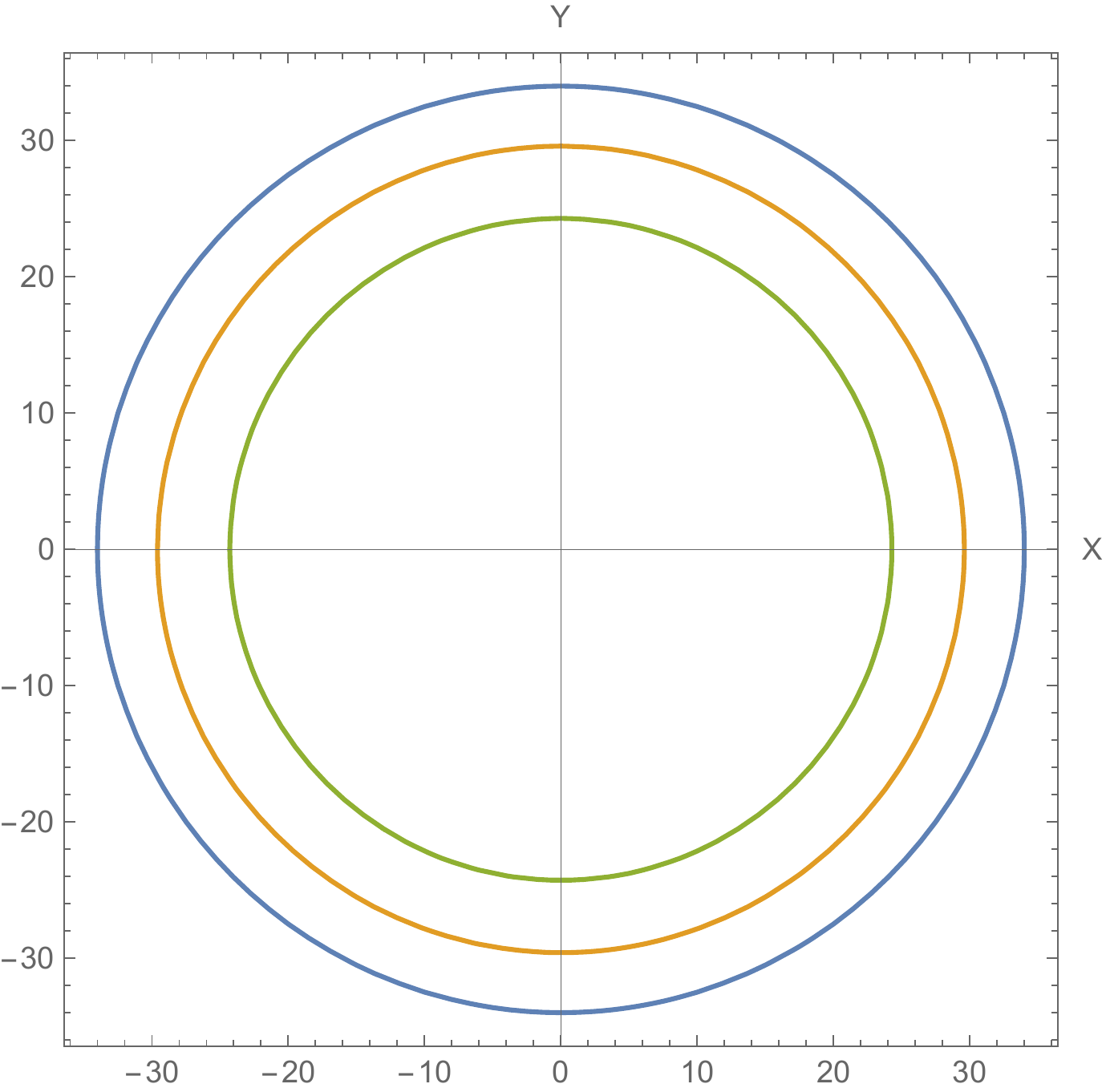}
\caption{Plot of the shadow contour with varying $\xi$, with $\beta=12$ and  $M=1$. 
The blue curve has $\xi=7.1$, the yellow $\xi=6.1$ and the green has $\xi=5.1$.}
\label{fig:sc with xi}
\end{minipage}
\end{figure}

In this case, as the metric represents an acoustic black hole with a single horizon, the profile of the  effective 
potential and the structure of the  related photon spheres are standard, and discussed in the literature widely. 
We will not go into the details here, and only present a plot of the effective potential in  
Fig. \ref{fig:CASE2} for completeness. Note again that the strong bending observables are also different from that of the 
above cases (1.1), (1.2) and in principle can serve as an important distinguishing feature between these two 
classes of geometries.

Similar scenarios occurs for case-3 and case-4 of previous section  even though multiple horizons are present in these cases.

Finally, we consider the shadow contour in our acoustic metric. 
The shadow of a typical black hole (or any compact object) for a faraway observer consists of the 
projection of the photon sphere to the observer's plane. For a non rotating metric, this reduces to 
finding two celestial coordinates $X, Y$ that describes the outline of the shadow by following 
the constraint relation $X^2+Y^2=r_{sh}^2$. Where, $r_{sh}$ is the shadow radius, which 
for an observer at infinity can be written down as $r_{sh}=\sqrt{\frac{\textbf{R}(r)^2}{G_{tt}}}\Big|_{r=r_{ph}}$. 
For our acoustic metric, in an entirely similar manner, we can define the shadow for motion of sound waves, 
which is a region for a ``listener at infinity'' that will be void of any sound. 

In Fig. \ref{fig:sc with xi}, we have plotted the shadow contour for various parametric values. 
Interestingly, the shadow radius increases with $\xi$ which characterises the medium in which the sound is 
propagating. And we can see that an increasing area of the listener's plane will be soundless with increasing 
$\xi$. In this figure, we have drawn the shadow contours for a value of $\beta$ in the wormhole branch for 
different values of $\xi$.

\section{Conclusions}
\label{sec6}

The traditional construction of the acoustic metrics for sound motion
in a fluid with a flat background (Minkowski space-time) has recently been
extended to include more general curved background space-time metrics. This
opens up a new and interesting direction to study sound motion in a
realistic black hole (or compact object) background immersed in some
fluid. In this paper, we have taken a step further in this direction to
include a general background metric into consideration, namely the
recently proposed Simpson-Visser black-bounce metric, which interpolates
between a regular black hole and a wormhole for different values of a
single parameter. 

Considering sound motion in a fluid
in this geometry, we have elucidated the resulting acoustic metric in
detail. We have shown that, the acoustic horizon may or may not be present
even in the absence of the optical event horizon depending on the
interplay between the parameters in question. As a result, sound can be trapped
(from the outside observer) in two ways, namely that the first one is  due solely to the
acoustic horizon present in the geometry, and the second one is due to the
optical event horizon itself, as the optical event horizon is  also an
acoustic horizon. The converse is however not true, and as a
result we can not have an acoustic wormhole in a black hole background. 
In summary, we have constructed a novel acoustic
metric which, depending on the parameter region, can broadly represent: 
(1) an acoustic wormhole and optical wormhole (WH-WH) geometry, (2)
acoustic black hole (with a single horizon or multiple horizons) and 
optical wormhole (BH-WH) geometry, (3) acoustic black hole and  optical
black hole (BH-BH) geometry.
This can be of importance from a experimental point of view, where
acoustic horizon, in the absence of a optical horizon, may be more easy
to simulate.

To elucidate the observational signatures of this metric, and to contrast how
this family is different from that of those that have appeared in the literature
(all of these are in black hole background, i.e., in our notation, of the BH-BH type), we have explored
one of the standard ways to distinguish a BH from a WH, namely the shadow
structure of the given metric. The special feature for light motion shown
by most wormhole metrics is the presence of a photon ring (and sphere) due to
light bending at the throat itself - the  sonic analogy of this fact
is used here to show how a similar feature for sound can produce a
\textit{sonic ring} due to the acoustic wormhole throat, in
addition to the known sonic rings due to the acoustic horizons. As a
result, sonic rings can still be produced even in the absence of any
acoustic horizons, which can have potential experimental realisation. In
this paper, we have taken a spherically symmetric background to describe
the acoustic metric. Slowly rotating backgrounds have been recently
discussed in the work of \cite{viera}, and a generalisation of our results to rotating backgrounds
and the observational signature of these should be an important
direction for future work.

\end{document}